**PRISM: Patient Response Identifiers for Stratified Medicine**

Thomas O. Jemielita[1] and Devan V. Mehrotra[1]

**Affiliation:**

[1]Biostatistics and Research Decision Sciences, Merck & Co., Inc. Kenilworth, NJ 07033, USA

**Abstract**

Pharmaceutical companies continue to seek innovative ways to explore whether a drug under development is likely to be suitable for all or only an identifiable stratum of patients in the target population. The sooner this can be done during the clinical development process, the better it is for the company, and downstream for prescribers, payers, and most importantly, for patients. To help enable this vision of stratified medicine, we describe a powerful statistical framework, Patient Response Identifiers for Stratified Medicine (PRISM), for the discovery of potential predictors of drug response and associated subgroups using machine learning tools. PRISM is highly flexible and can have many "configurations", allowing the incorporation of complementary models or tools for a variety of outcomes and settings. One promising PRISM configuration is to use the observed outcomes for subgroup identification, while using counterfactual within-patient predicted treatment differences for subgroup-specific treatment estimates and associated interpretation. This separates the "subgroup-identification" from the "decision-making" and, to facilitate clinical design planning, is a simple way to obtain unbiased treatment effect sizes in the discovered subgroups. Simulation results, along with data from a real clinical trial are used to illustrate the utility of the proposed PRISM framework.

## 1 Introduction

Most medical treatments are designed and approved for the "average patient" in a one-size-fits-all approach. The downside of this approach is that if the "average patient" within a broad population benefits from a study treatment, it is possible there are patients who receive no or limited benefit as well as patients who receive enhanced benefit. The overall goal of precision medicine, which focuses on the individual level, and stratified medicine, which focuses on finding interpretable subgroups, is to understand the underlying cause of disease, evaluate treatment response heterogeneity, and determine what types of patients respond or do not respond to a given treatment[1]. While this can be assessed post-hoc if the drug is shown to be beneficial, on average, in the overall population, this approach misses the opportunity to leverage information on treatment response heterogeneity for improving both subsequent trial design and other decision making. Further, along with recent advances in technology and data availability, rapid identification of subgroups has the potential to improve patient care and accelerate drug development[2].

The focus on this work is more aligned with stratified medicine, where we aim to identify subgroups of patients based on treatment response heterogeneity. For example, consider a Phase

II study where an experimental drug is compared to a control. Suppose the strategy is to test for efficacy only in the overall population and, if successful, a confirmatory Phase III trial is initiated. Three potential scenarios could occur: (1) The drug is truly efficacious for all patients and this strategy will be successful no matter what, (2) The drug proves efficacious overall, but the treatment effect is diluted by patients who receive none or limited benefit, or (3) The drug is not efficacious overall but certain patients do receive clinically relevant benefit. Although all patients benefit in Scenario (1), knowledge of treatment response heterogeneity can improve design planning and analysis of the subsequent Phase III trial. In Scenario (2), understanding that some patients do not respond to the treatment can also improve the trial design. For example, suppose that patients with a certain genotype seem to benefit but those without do not. Genotype could then be a stratification factor in Phase III. Alternatively, if there is strong enough evidence, such as biological rationale or other supporting information, the Phase III study could focus on patients with the given genotype. In Scenario (3) where the drug is not efficacious overall, the decision may be to abandon the drug altogether even though there may be patients that truly benefit from the study drug. In any these scenarios, drug development is improved by understanding both the overall treatment effect along with subgroup specific treatment effects.

Overall, there are four broad frameworks for subgroup identification: global outcome modeling, global treatment effect modeling, optimal treatment regimes, and local modeling[3]. Global outcome modeling fits a model (or treatment specific models) that aims to capture treatment-by-covariate interactions. The resulting predictions can then be used to form subgroups. Here, ensemble models and penalized regression are commonly employed such as the non-parametric Virtual Twins and FindIT methods[4,5]. Global treatment effect modeling aims to partition the covariate space by directly targeting the treatment interactions. For example, the interaction tree method extends the classic CART method[6] by including a treatment-by-split interaction and thus searches for partitions related to treatment response heterogeneity[7-9]. Other tree-based methods include GUIDE[10] and model-based partitioning (MOB)[11-13]. Both allow for the flexibility to look for partitions corresponding to prognostic covariates (associated with the outcome irrespective of the treatment assignment), predictive covariates (associated with the outcome conditional on treatment), or both. In contrast to the previous frameworks, optimal treatment regimes focus on identifying the optimal treatment for each patient, rather than finding the best type of patients for each treatment. This approach often includes global outcome modeling[14,15]. Lastly, local modeling searches for regions of the covariate space with improved treatment effects[3]. This avoids needing to estimate an outcome function over the entire covariate space and instead focuses in on specific covariate space regions. Examples include SIDES, SIDESCREEN, and PRIM[16-18]. Lastly, it is worth mentioning the Bayesian Credible Subgroup methodology[19]. Here, the idea is to find covariate regions where all members simultaneously have a treatment effect exceed some pre-specified threshold.

Our proposed method, PRISM (Patient Response Identifiers for Stratified Medicine), is a general-purpose subgroup identification approach that can incorporate a wide variety of machine learning tools and subgroup identification models. As noted above, there are many approaches to subgroup identification and a key motivation for PRISM was to consolidate these various approaches into a unified framework. This facilitates the construction of PRISM "configurations" that can combine complementary methods or models. The key output is rule-based subgroups (if any) and, for both the overall and discovered subgroups, point-estimates and variability metrics (standard errors (SE), confidence intervals (CIs), probability statements) for decision making. Obtaining unbiased estimates is especially crucial for drug development since if a subgroup is identified and moved forward for further testing, biased estimated effect sizes may lead to incorrect estimates of the required sample size and thus compromise the power of any future study.

In the next Section, we describe a motivating example involving a clinical trial from the anti-infectives domain. In-depth details of PRISM are provided in Section 3. The operating characteristics of PRISM are examined using a simulation study in Section 4 followed by a PRISM-based analysis of the motivating clinical trial data in Section 5. We conclude with a summary of results and a discussion of future directions.

## 2    Clinical Trial Example: Clostridium Difficile Recurrence

The data for our motivating example come from a phase III, double-blind, randomized, placebo-controlled clinical trial of an experimental drug (bezlotoxumab) in patients being treated for Clostridium Difficile (CDIFF)[20] infection (CDI). CDIFF is the most common cause of infectious diarrhea in hospitalized patients and is estimated to cause nearly half a million cases in the US alone[21]. The infection is typically treated with antibiotics; despite the resulting "clinical cure" for most patients, many experience a recurrence of CDIFF infection within a relatively short time window.  The phase III trial was designed to assess whether bezlotoxumab, relative to placebo, reduces the risk of CDI recurrence, the latter defined as the development of a new episode of diarrhea (3 or more loose stools in 24 or fewer hours) and a positive local or central lab stool test for toxigenic CDIFF following clinical cure of the initial CDI episode. There were two bezlotoxumab containing arms; for illustration, both are combined into a single arm and referred to as the bezlotoxumab arm. Of interest here is the CDI recurrence risk difference between the bezlotoxumab and placebo arms with a clinically meaningful difference of -0.10, i.e., a 10 percentage point lower CDI recurrence risk for bezlotoxumab versus placebo. Five baseline variables were pre-specified in the protocol as risk factors for CDI recurrence: prior history of CDI within past 6 months, immune compromised (Yes or No), CDI severity (Yes or No), prior history of CDI (ever, Yes or No), hypervirulent strain (Yes vs No) and age (<65 vs >=65).  An additional variable was subsequently identified in a genome-wide association study (GWAS): SNP+ (yes if patient had one or more copies of the minor allele for SNP rs251613, and no otherwise). All six variables will be used to illustrate the PRISM approach in the remainder of this article, and the analysis set is restricted to patients who had provided consent for the GWAS, resulting in sample sizes of 240 and 461 for placebo and bezlotoxumab, respectively.

The forest plot in Figure 1 shows the observed CDI recurrence rates for each treatment and for the corresponding treatment difference separately for subgroups defined by the aforementioned six variables. Consistent with the original findings of the study, the presence of any of the following conditions resulted in an increased treatment effect: patients with Prior CDI (6 mon), immune compromised (yes), CDI severity (yes), SNP+, Age >=65, and Hypervirulent strain (yes). However, it isn't exactly clear how to interpret these results. For example, what if patient had Prior CDI history but was age <65? What if a patient was SNP- but had prior CDI history? While the univariate forest plot is informative, interpretation is difficult and response to treatment is likely based on multiple baseline variables. This is where PRISM comes in.

**Figure 1:** Forest Plot, CDIFF Example

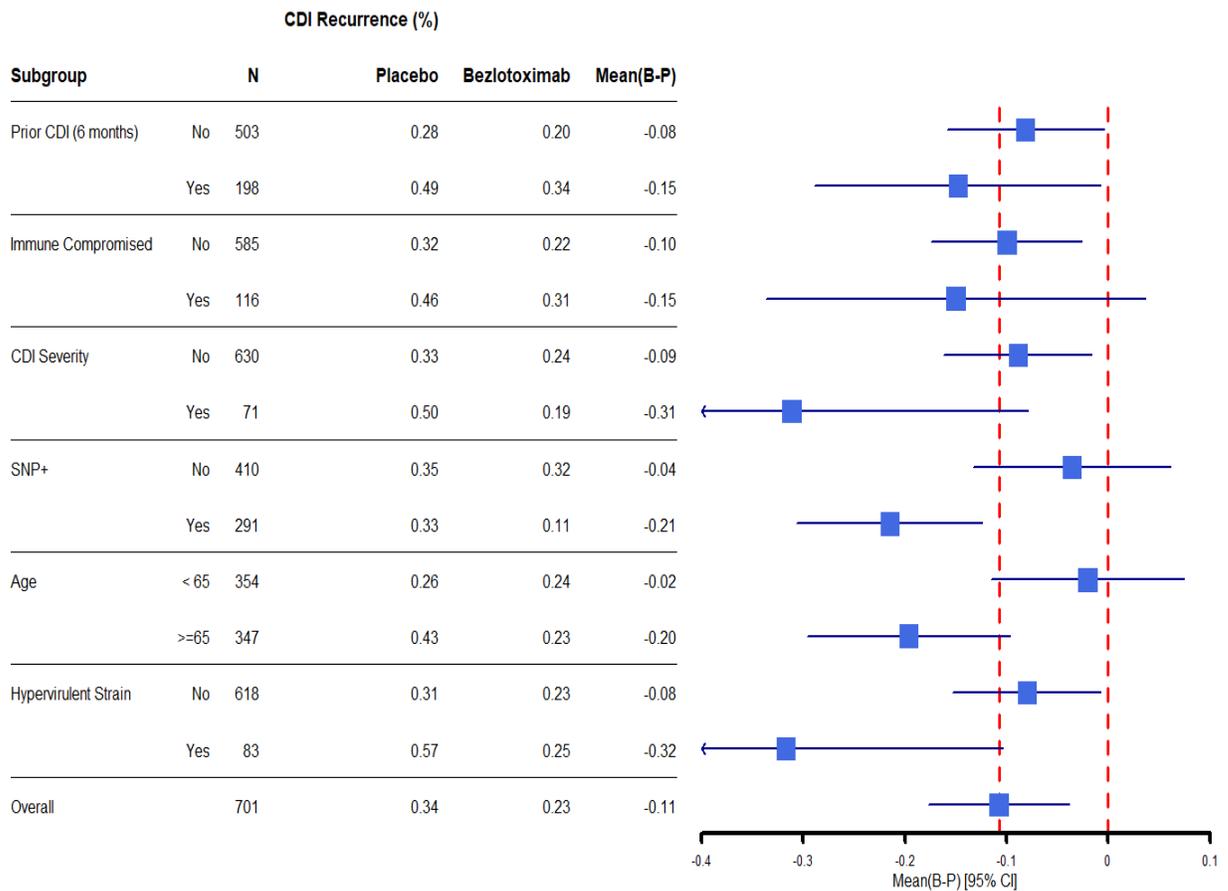

## 3    Methods

The PRISM framework offers a roadmap for subgroup identification and subsequent decision making. PRISM consists of five steps which are summarized in Figure 2. The first step is to determine the question(s) or estimand(s) of interest. For example, what is the average treatment difference between the two treatments being compared? The second step filters out "noise" variables from a larger candidate set of baseline variables, with a goal of retaining variables that are plausibly related to the outcome for at least one treatment. The third step outputs "Patient-Level Estimates" (PLEs). PLEs are patient-level counterfactual quantities that relate to the question of interest. In our running example, this corresponds to the patient-level treatment effect, or the expected difference in response if the patient had received both treatments. Understanding patient-level treatment effects is a key aspect of both stratified and precision medicine. The fourth step determines rule-based interpretable subgroups. This can be based off the observed outcomes, the PLEs, or both. The final step involves parameter estimation and decision making both for the overall population and for the identified subgroups. To summarize, PRISM provides insight at the overall population level (is there a treatment effect, on average?) and at the patient or subgroup level (are there heterogeneous treatment effects?).

PRISM can be implemented through the R package "StratifiedMedicine." One novel feature of PRISM is that users can input their own methods or functions at each step of the algorithm. This flexibility allows for easy research-experimentation and timely incorporation of state-of-the-art modeling techniques. We now describe PRISM in detail.

## Figure 2: PRISM (Patient Response Identifiers for Stratified Medicine)

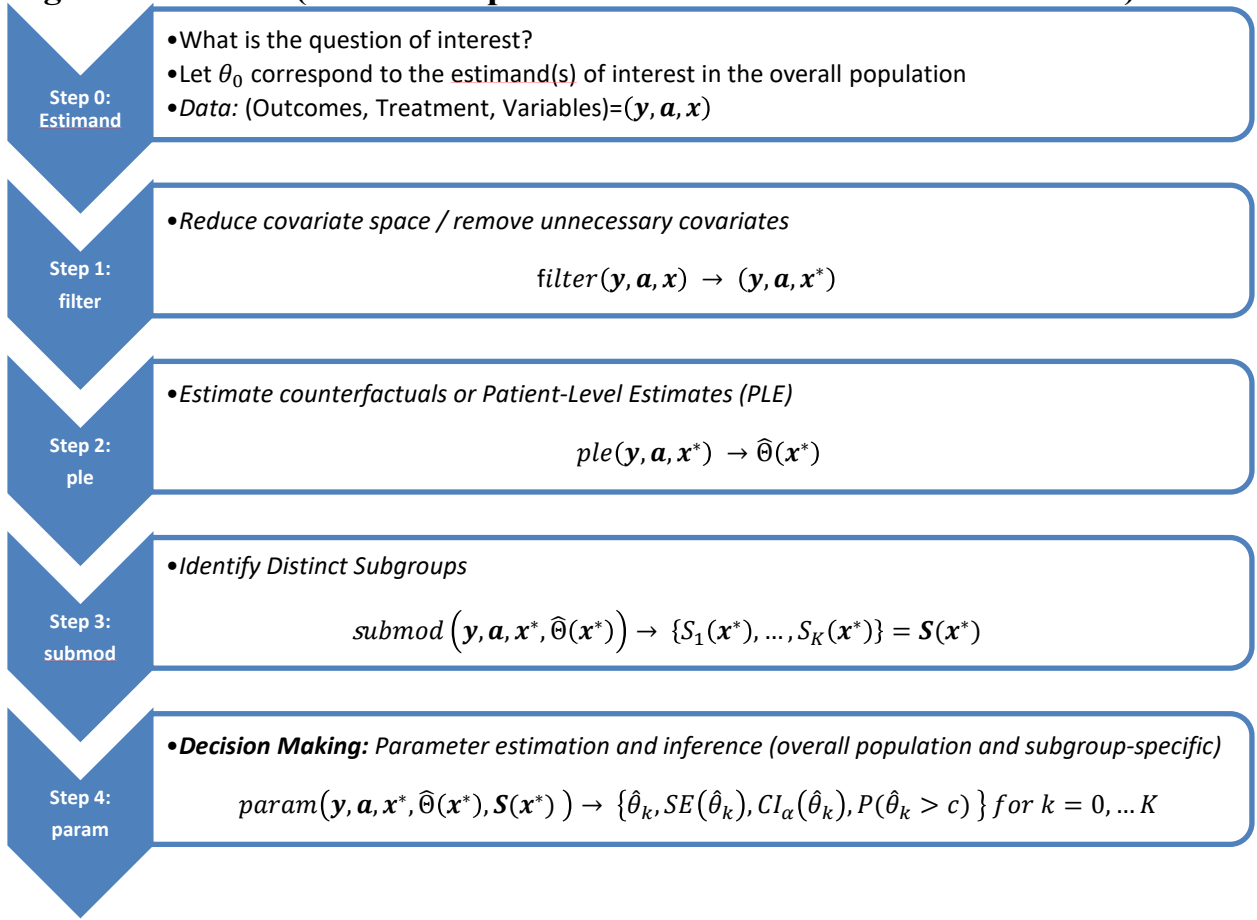

**Step 0: Estimand**
- What is the question of interest?
- Let $\theta_0$ correspond to the estimand(s) of interest in the overall population
- *Data:* (Outcomes, Treatment, Variables)=$(\boldsymbol{y}, \boldsymbol{a}, \boldsymbol{x})$

**Step 1: filter**
- *Reduce covariate space / remove unnecessary covariates*

$$filter(\boldsymbol{y}, \boldsymbol{a}, \boldsymbol{x}) \to (\boldsymbol{y}, \boldsymbol{a}, \boldsymbol{x}^*)$$

**Step 2: ple**
- *Estimate counterfactuals or Patient-Level Estimates (PLE)*

$$ple(\boldsymbol{y}, \boldsymbol{a}, \boldsymbol{x}^*) \to \widehat{\Theta}(\boldsymbol{x}^*)$$

**Step 3: submod**
- *Identify Distinct Subgroups*

$$submod\left(\boldsymbol{y}, \boldsymbol{a}, \boldsymbol{x}^*, \widehat{\Theta}(\boldsymbol{x}^*)\right) \to \{S_1(\boldsymbol{x}^*), \dots, S_K(\boldsymbol{x}^*)\} = \boldsymbol{S}(\boldsymbol{x}^*)$$

**Step 4: param**
- ***Decision Making:*** *Parameter estimation and inference (overall population and subgroup-specific)*

$$param(\boldsymbol{y}, \boldsymbol{a}, \boldsymbol{x}^*, \widehat{\Theta}(\boldsymbol{x}^*), \boldsymbol{S}(\boldsymbol{x}^*)) \to \{\hat{\theta}_k, SE(\hat{\theta}_k), CI_\alpha(\hat{\theta}_k), P(\hat{\theta}_k > c)\} \text{ for } k = 0, \dots K$$

For $i = 1, \ldots, n$ patients, let $(\boldsymbol{Y}_i, A_i, \boldsymbol{X}_i)$ represent the patient level random variable data structure where $\boldsymbol{Y}_i$ is $J$-dimensional vector of patient outcomes (binary, continuous, ordinal, time-to-event, etc), $A_i$ is a treatment variable, and $\boldsymbol{X}_i = (X_{i1}, \ldots, X_{ip})^T$ is a p-dimensional covariate vector. The vector of outcomes $\boldsymbol{Y}_i$ could be a vector of longitudinal measurements, a combination of safety and efficacy outcomes (benefit-risk), or a single efficacy measure. The treatment variable could correspond to a binary treatment variable where 1 corresponds to the test/experimental drug and 0 corresponds to the control, or perhaps multiple treatments or dose levels. The observed values of the random variables are denoted as $(\boldsymbol{y}_i, a_i, \boldsymbol{x}_i)$. Lastly, define the observed "stacked" data as $(\boldsymbol{y}, \boldsymbol{a}, \boldsymbol{x})$ where $\boldsymbol{y}$ is a $n$ by $j$ matrix, $\mathbf{a}$ is a n-length vector, and $\boldsymbol{x}$ is a $n$ by $p$ matrix. Next, each step of PRISM in described in detail. For context, the running example is a randomized control trial (RCT) where patients are randomized 1:1 to either the test drug or the control and the outcome of interest is continuous (e.g., tumor size change) or binary (e.g., CDI recurrence).

**Step 0: Estimand**

The first step is to determine the estimand(s) or question of interest(s). Let $\theta_0$ correspond to the estimand(s) of interest in the overall population with corresponding estimate $\hat{\theta}_0$. For ease of notation, $\theta_0$ could refer to multiple estimands or a single estimand. For example, for a continuous efficacy outcome and binary treatment, the estimand of interest is often the average treatment effect (ATE), $\theta_0 = E(Y|A = 1) - E(Y|A = 0)$. The ATE estimate, $\hat{\theta}_0 = \hat{E}(Y|A = 1) - \hat{E}(Y|A = 0)$, could then be obtained through ordinary least squares (OLS) adjusted for some number of prognostic and/or predictive covariates. Alternatively, there may be multiple estimands of interest:

$$\theta_0 = [E(Y|A = 0), E(Y|A = 1), E(Y|A = 1) - E(Y|A = 0)].$$

**Step 1: Filter**

The second step is to filter variables that are deemed unrelated to the outcomes for either treatment. While there are trade-offs for any filtering approach, removing "noise" variables that are completely unrelated to the outcomes and/or treatments will enhance the performance of any subgroup identification method. Formally:

$$filter(\boldsymbol{y}, \boldsymbol{a}, \boldsymbol{x}) \ \rightarrow \ (\boldsymbol{y}, \boldsymbol{a}, \boldsymbol{x}^*)$$

where $filter(.)$ is some model based on the observed data and outputs a potentially smaller covariate space $\boldsymbol{x}^*$ with dimension $n \times q$ where $q \leq p$. Examples of filtering tools include univariate p-values, elastic net[22], and random forest variable importance with confidence intervals[23].

**Step 2: Patient-Level Estimate (PLE)**

Define the *PLE* model as:

$$ple(\boldsymbol{y}, \boldsymbol{a}, \boldsymbol{x}^*) \rightarrow \left( \widehat{\Theta}(\boldsymbol{x}^*_1), \dots, \widehat{\Theta}(\boldsymbol{x}^*_i), \dots, \widehat{\Theta}(\boldsymbol{x}^*_n) \right) = \widehat{\Theta}(\boldsymbol{x}^*)$$

where $ple(.)$ is some model based on the observed data which outputs a matrix of PLEs where $\widehat{\Theta}(\boldsymbol{x}^*_i)$ contains the patient-level estimates of $\theta_0$. For the continuous outcome and binary treatment example, the PLE is often referred to as the estimated individual treatment effect (ITE). Let $\mu(a, \boldsymbol{x}^*) = E(Y|A = a, X^* = \boldsymbol{x}^*)$ denote the outcome regression model with estimates $\hat{\mu}(a, \boldsymbol{x}^*)$. Then:

$$\widehat{\Theta}(\boldsymbol{x}^*_i) = \hat{\mu}(a = 1, \boldsymbol{x}^*_i) - \hat{\mu}(a = 1, \boldsymbol{x}^*_i)$$

PLEs are informative of the underlying patient heterogeneity and subgroups can be directly identified by partitioning the PLEs. For the continuous outcome example, note that $\hat{\mu}(a, \boldsymbol{x}^*_i)$ corresponds to the estimated expected response as a function of the vector $\boldsymbol{X}^*_i$ and treatment assignment $A_i$ evaluated at $\boldsymbol{x}^*_i$ and $a$. In parallel arm trials, patients do not receive both treatments and PLEs must be estimated from the data. For the continuous outcome example, possible PLE models include linear models with explicit treatment by covariate interactions (e.g., $Y \sim \beta X + \alpha A + \gamma XA$), fitting random forest models for each treatment group separately[4,24], the related causal forest method[25], and BART[26].

**Step 3: Subgroup Identification (submod)**

The next step is to find rule-based and interpretable subgroups. Define the *subgroup model* as:

$$submod \left( \boldsymbol{y}, \boldsymbol{a}, \boldsymbol{x}^*, \widehat{\Theta}(\boldsymbol{x}^*) \right) \rightarrow \{S_1(\boldsymbol{x}^*), \dots, S_K(\boldsymbol{x}^*)\} = \boldsymbol{S}(\boldsymbol{x}^*)$$

where $submod(.)$ maps the observed data and/or PLEs to $k = 1, \dots, K$ distinct subgroups and $S_k(\boldsymbol{x}^*)$ defines some mapping of the covariate space for subgroup $k$. Subgroups can be defined by one binary variable, $\{S_1(\boldsymbol{x}^*): [X_1 = 0], \ S_2(\boldsymbol{x}^*): [X_1 = 1] \}$, as well as by multiple variables, for example $\{S_1(\boldsymbol{x}^*): [X_l < c, \ X_j = 1], \ S_2(\boldsymbol{x}^*): [X_l \geq c, X_j = 1], \ S_3(\boldsymbol{x}^*): [X_j = 0] \}$ where subgroups are defined by $X_j = 0$ and within $X_j{=}1$ by some cutoff $c$ for continuous variable $X_l$. Examples of subgroup models include VirtualTwins[4], CART[6], CTREE[11], MOB[12], SIDES[16], interaction trees[8], and GUIDE[27].

**Step 4: Decision Making (param)**

The final step is parameter estimation and inference within the discovered subgroups and the overall population. Importantly, the point estimates in the overall and subgroup-specific populations should be unbiased, i.e. $E(\hat{\theta}_k) = \theta_k$ and $E(\hat{\theta}_0) = \theta_0$. The standard error of the estimate $SE(\widehat{\theta_k})$, confidence intervals with nominal level $1-\alpha$, $CI_\alpha(\hat{\theta}_k)$, and probability statements, for example $P(\hat{\theta}_k > c)$ where $c$ is some threshold, are also needed for decision making. Define the *parameter* model as:

$$param\big(\boldsymbol{y}, \boldsymbol{a}, \boldsymbol{x}^*, \widehat{\boldsymbol{\theta}}(\boldsymbol{x}^*), \boldsymbol{S}(\boldsymbol{x}^*)\big) \rightarrow \big\{\hat{\theta}_k, SE(\hat{\theta}_k), CI_\alpha(\hat{\theta}_k), P(\hat{\theta}_k > c)\big\} \, for \, k = 0,1, \dots K$$

where $param(.)$ is some function that outputs subgroup-specific, as well as in the overall population at $k = 0$, point-estimates, SEs, CIs, and probability statements based on the identified subgroup rules and the observed data and/or PLEs. These metrics can also come directly from the subgroup model (Step 3). For the continuous outcome and binary treatment example, one approach is to fit separate linear regression models within each identified subgroup. Conceptually, this is what MOB does[13]. Lastly, in line with the multiple estimand setting (i.e. multiple treatment effects, or a survival curve), the $param(.)$ function could instead output multiple estimates for each discovered subgroup and the overall population.

For the most part, subgroup-specific parameter estimation and subsequent inference using the observed data has two main problems: (1) $\hat{\theta}_k$ can be biased since the subgroup identification model and parameter model both use the same data, and (2) SEs, CIs, and probability statements ignore the inherent subgroup selection (as well as the estimation of the PLEs and filtering) and inference through these metrics is generally not valid. Non-parametric bootstrapping, which repeats PRISM Steps 1-4 across resamples of the data, is one general solution to these problems (Details in Appendix). Another semi-parametric/Bayesian approach, which doesn't require resampling, is also discussed below. Both performed similarly in the simulations and real data example.

### 3.1 Parameter Estimation: PLE

The PLE approach obtains subgroup-specific point estimates by averaging the PLEs within each subgroup and in the overall population:

$$\tilde{\theta}_k = \frac{1}{n_k} \sum_{i \in S_k}^{n_k} \hat{\theta}(\boldsymbol{x}^*_i) \quad (1)$$

where $n_k$ is the sample size of subgroup $k$ with $n_0 = n$ (overall sample size) and $i \in S_k$ corresponds to the patients included in subgroup $k$ for $k = 0,1, \dots, K$. Next, define the patient specific "pseudo-outcomes" as:

$$y^*_i = \left(\frac{a_i\big(y_i - \hat{\mu}(a=1, \boldsymbol{x}^*_i)\big)}{\hat{\pi}(\boldsymbol{x}^*_i)} - \frac{(1-a_i)\big(y_i - \hat{\mu}(a=0, \boldsymbol{x}^*_i)\big)}{1 - \hat{\pi}(\boldsymbol{x}^*_i)}\right)$$
$$+ \big(\hat{\mu}(a=1, \boldsymbol{x}^*_i) - \hat{\mu}(a=0, \boldsymbol{x}^*_i)\big)$$

where $\hat{\pi}(\boldsymbol{x}^*_i) = \hat{P}(a=1|\boldsymbol{X}^*_i = \boldsymbol{x}^*_i)$ is the patient-level estimate of receiving treatment $a$=1. Note that this relates to the so-called "double robust" estimators[28]. The main difference is that we only use the model fits (the second term above). Compared to only using the model fits for parameter estimation, taking the average of the pseudo-outcomes for parameter estimation results in higher bias when using MOB to find subgroups. This is because the pseudo-outcomes explicitly use the observed outcomes, which are also used to find the subgroups, for parameter estimation. Importantly, if the estimated probabilities or the estimated expectations are consistent, then:

$$E(y^*_i) = E(Y|A_i = 1, \boldsymbol{X}^*_i = \boldsymbol{x}^*_i) - E(Y|A_i = 0, \boldsymbol{X}^*_i = \boldsymbol{x}^*_i)$$

$$E\left(n_k^{-1} \sum_{i \in S_k}^{n_k} y^*_i\right) = E\left(Y|A_i = 1, \boldsymbol{x}^* \in S_k(\boldsymbol{x}^*)\right) - E\left(Y|A_i = 0, \boldsymbol{x}^* \in S_k(\boldsymbol{x}^*)\right) = \theta_k$$

For a randomized clinical trial, the treatment assignment mechanism is known and $\hat{P}(a = 1|\boldsymbol{X}^*_i = \boldsymbol{x}^*_i)$ can be replaced with the marginal estimate $\hat{P}(a = 1)$. The standard error $SE(\hat{\theta}_k)$ is then calculated as:

$$SE(\tilde{\theta}_k) = \sqrt{n_k^{-2} \sum_{i \in S_k}^{n_k} (y^*_i - \tilde{\theta}_k)^2} \quad (2)$$

where $SE(\tilde{\theta}_k)^2 = Var(\tilde{\theta}_k)$. CIs can then be formed using Z or t-intervals, for example, $[\tilde{\theta}_k \pm t_{1-\alpha/2, n_k-1} SE(\tilde{\theta}_k)]$.

## 3.2    Parameter Estimation: Bayesian

Based on our simulation studies and in line with intuitive expectations, the PLE parameter estimation approach yields relatively unbiased point-estimates with valid coverage. Based on these initial PLE parameter estimates ($\tilde{\theta}_k$), we propose a simple Bayesian approach to obtain a posterior distribution and thus calculate meaningful probability statements. For each identified subgroup, the posterior distribution of $p(\tilde{\theta}_k|data)$ is estimated by assuming a normal prior with mean $\hat{\theta}_0$ and variance $\gamma * Var(\hat{\theta}_0)$ where $\gamma$ is some scaling factor. In this research, we set $\gamma = n$ such that the prior is uninformative. Following standard normal theory:

$$(\tilde{\theta}_k|data) \sim N\left(\hat{\theta}_k, Var(\hat{\theta}_k)\right) \quad (3)$$

$$Var(\hat{\theta}_k) = \left(1/(\gamma * Var(\hat{\theta}_0)) + 1/Var(\tilde{\theta}_k)\right)^{-1} \quad (4)$$

$$\hat{\theta}_k = Var(\hat{\theta}_k) * \left(\hat{\theta}_0/(\gamma * Var(\hat{\theta}_0)) + \tilde{\theta}_k/Var(\tilde{\theta}_k)\right) \quad (5)$$

Frequentist CIs can be estimated by taking the $\alpha/2$ and $1 - \alpha/2$ quantiles of the posterior distribution. Similarly, probability statements are directly obtained from the posterior. Lastly, while we assumed a normal posterior (with a normal prior), this could be easily be extended to different distributions (e.g., binomial).

## 4   Simulation Study

For this simulation study and the real data example, we describe three example PRISM configurations: PRISM(A), PRISM(B), and MOB. These are applicable to both the real data example and the simulations and details can be found in Table 1. MOB is a tree-based method that partitions the data based on prognostic and/or predictive variables. For the continuous outcome simulations, we use OLS-based MOB (lmtree); for the binary simulations and real-data example, since the main interest is the risk difference, we use GLM-based MOB (glmtree; binomial with identity link). OLS/GLM (binomial with identity link) models are then fit within each subgroup. PRISM(A) and PRISM(B) both include an elastic net filter, PLE through the "counterfactual forest" (random forest models), and parameter estimation with the PLEs or counterfactual estimates (with Bayesian update). For elastic net, the treatment variable is not included and thus focuses on finding prognostic variables. The key difference between PRISM(A) and PRISM(B) is the subgroup model: PRISM(A) identifies subgroups using the observed outcome and MOB, while PRISM(B) identifies subgroups using the counterfactual PLEs and CTREE.

**Table 1: PRISM Configurations: Continuous or Binary Outcome with Binary Treatment**

| PRISM Steps | Configurations | Methods | | |
|---|---|---|---|---|
| | | **MOB** | **PRISM(A)** | **PRISM(B)** |
| **Filter** | **Elastic Net (ENET)** $$ENET(\boldsymbol{y}, \boldsymbol{x}) \rightarrow (\boldsymbol{y}, \boldsymbol{a}, \boldsymbol{x}^*)$$ *R package: glmnet (default settings, alpha=0.5)* | | Yes | Yes |
| **PLE** | **Counterfactual Forest (CF)** $$RF_1(1, \boldsymbol{x}^*) - RF_0(0, \boldsymbol{x}^*) \rightarrow \widehat{\boldsymbol{\theta}}(\boldsymbol{x}^*)$$ *Details:* Fit RF models within each treatment group. *R package: ranger (default regression forest, min.node.size=10% of total sample size)* | | Yes | Yes |
| **Submod** | **MOB** $$MOB(\boldsymbol{y}, a, \boldsymbol{x}^*) \rightarrow \boldsymbol{S}(\boldsymbol{x}^*)$$ *R package: partykit* *Continuous Outcome: lmtree (Gaussian)* *Binary Outcome: glmtree (Binomial with identity link)* **CTREE** $$CTREE(\widehat{\boldsymbol{\theta}}(\boldsymbol{x}^*), \boldsymbol{x}^*) \rightarrow \boldsymbol{S}(\boldsymbol{x}^*)$$ *R package: partykit / ctree* *Details***:** For MOB/CTREE, default settings with alpha=0.10, max.depth=4, minimum node size set to *10% of total sample size*. | Yes | Yes<br><br>Yes | Yes |
| **Param** | **OLS/GLM:** Fit $y \sim a$ within each subgroup **PLE:** Average PLEs within each subgroup. Apply Bayesian update and calculate Bayesian posterior. | Yes | Yes | Yes |

The purpose of the simulation study was to evaluate the performance of PRISM with respect to (1) How often were subgroups formed based on the true predictive covariates, (2) Impact of filtering as more noise is added into the baseline covariate space, (3) Bias, relative efficiency, and coverage of the treatment effect estimates at the subgroup level, and (4) Compared to standard practice (i.e. treat all patients if overall treatment effect p-value<0.05), how effective is PRISM at assigning the "right" treatment to patients?

For all simulations, a sample size of N=800 was generated with a 1:1 randomization ratio for the test drug (*A=1*) and control (*A=0*). The outcome was continuous (t-distribution) or binary (overall response rate of 30%) and there were, in truth, three covariates that were predictive (as well as prognostic) which defined subgroups with varying treatment effects. There were also 3 purely prognostic covariates that were correlated with the predictive covariates. In addition, there were either 6 or 56 noise variables. Two general subgroup scenarios were explored (1) Null setting where there is no treatment effect for any patient, and (2) Subgroup setting where there are in truth 4 subgroups. An "eight" subgroup scenario was also examined, but ultimately yielded similar results to the 4-subgroup scenario. In the Subgroup setting, while the overall treatment effect is non-zero (0.237 (continuous), 0.11 (binary)), 30% of the patients receive no benefit from the test drug versus the control. Subgroups were defined by a binary variable and two continuous variables. 1000 simulations were generated for each scenario. More details on the simulations can be found in the Appendix.

## 4.1   Filtering and Subgroup Identification: Variable Selection

These results focus on the Subgroup setting. See Figure 3. This plot shows the average percent correctly selected for predictive variables (3 in truth) and noise variables (6 or 56) that define the identified subgroups in PRISM(A)/PRISM(B) by filtering (No Filtering or ENET). As a reminder, PRISM(A) uses the observed outcomes and MOB while PRISM(B) uses the PLEs in CTREE for subgroup identification. Here we highlight key insights. In general, as expected, filtering improves the ability of each subgroup model to form subgroups based on the true predictive variables; this is especially true as more noise variables are added. For example, with 6 noise variables and a continuous outcome, PRISM(A) selects ~80% of the true predictive with or without filtering, but with 56 noise variables, PRISM(A) with or without filtering selects 75% and 68% of the true predictive variables respectively. Compared to PRISM(B) (with or without filtering), PRISM(A) tends to select more of the true predictive variables and less of the true noise variables. For example, with filtering in the Subgroup scenario with 56 noise variables and a binary outcome, PRISM(B) and PRISM(A) select 57% and 88% of the true predictive variables respectively and 5% (~3 noise variables) and 0.2% (<1 noise variables) of the true noise variables respectively. Lastly, while we don't present results for prognostic variables, PRISM(A) tended to select ~20-30% of the purely prognostic variables and PRISM(B) tended to select ~10-50% of the purely prognostic variables. In the Null setting, while there were no predictive variables, filtering also helped remove noise variables and select the true prognostic variables.

**Figure 3: Filtering and Subgroup Models: Variable Selection**

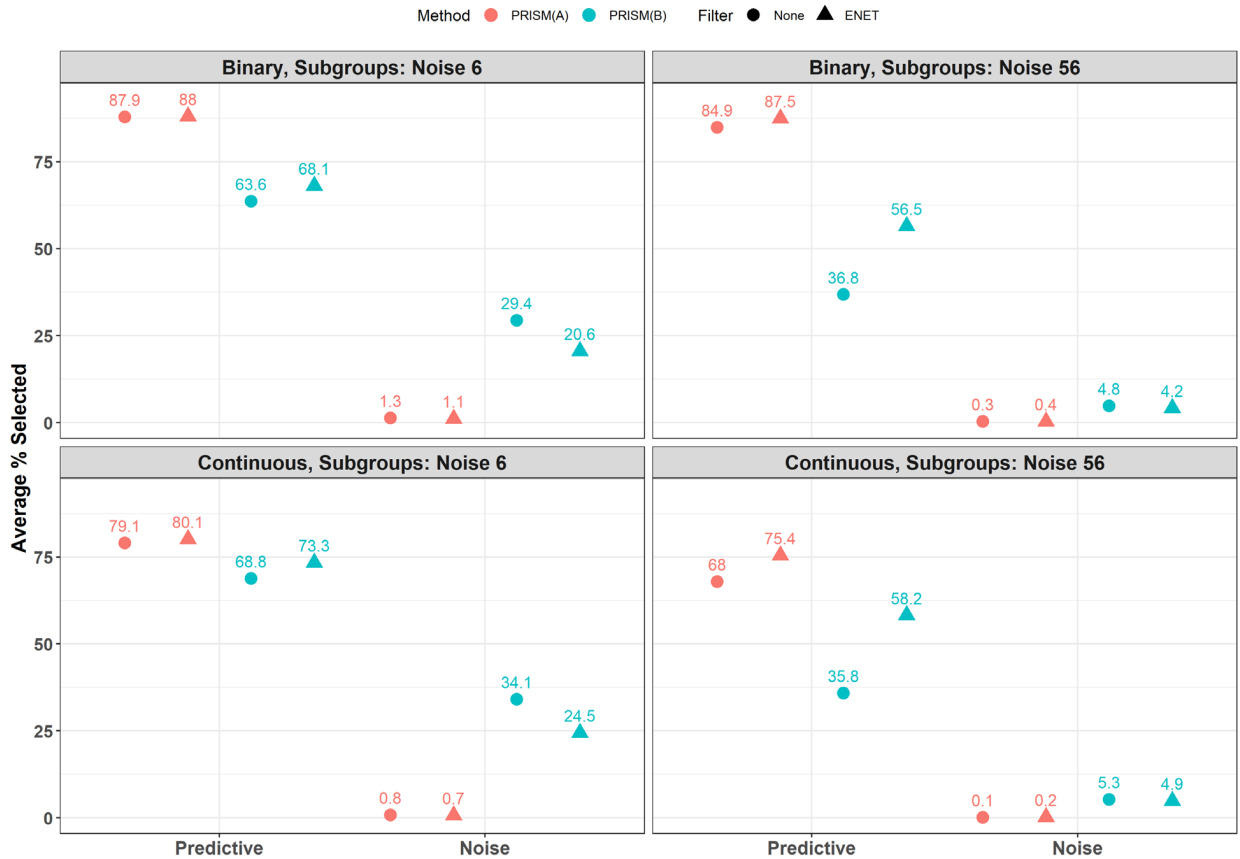

**Note:** This plot illustrates the ability of different subgroup models, with and without filtering, to detect the true predictive/prognostic variables (in truth 3) and to variables (6 or 56). Key points: (1) Filtering improves the ability of any subgroup model and (2) Compared to PRISM(B) (subgroups based on counterfactual estimates using CTREE), PRISM(A) (subgroups based on observed outcomes using MOB) tends to select more of the truly predictive/prognostic variables and less of the noise variables.

## 4.2    Bias, Relative, and Coverage

As a benchmark, we include the "oracle" method which, for continuous outcomes, fits OLS models within each of the true subgroups, and for binary outcomes, estimates the risk difference within each of the true subgroups based on Miettinen-Nurminen method[29]. This represents an ideal situation where we know exactly how many and which subgroups there are. For the Null setting, this corresponds to fitting a single model in the overall population. For these simulations, results were similar with 6 or 56 noise variables and these settings were combined. See Figure 4 for bias, relative efficiency, and coverage results. Starting with overall bias, note that all methods have approximately zero average bias. In terms of absolute bias, which avoids cancelling out large positive and large negative bias with subgroups, PRISM(A) yields the lowest absolute bias. For relative efficiency, values < 1 correspond to MOB having superior efficiency. In the Null setting, the "oracle" method by far performs the best; mainly because the "oracle" knows there are no subgroups. PRISM(A), relative to MOB and PRISM(B), also shows increased efficiency. In the Subgroup setting, PRISM(A) yields the highest efficiency. Lastly, for coverage, note that the dashed line corresponds to 93.6%, which is two standard errors (based on 1000 simulations) below the desired 95% coverage. Both the "oracle" and PRISM(A) consistently maintain 95% coverage at the subgroup level. PRISM(B) yields valid coverage in most scenarios, but slightly below the simulation error bound in the Continuous-Null scenario. MOB, which uses OLS/GLM for parameter estimation, does not reach the desired 95% coverage in any of the scenarios.

**Figure 4: Bias, Relative Efficiency, and Coverage**

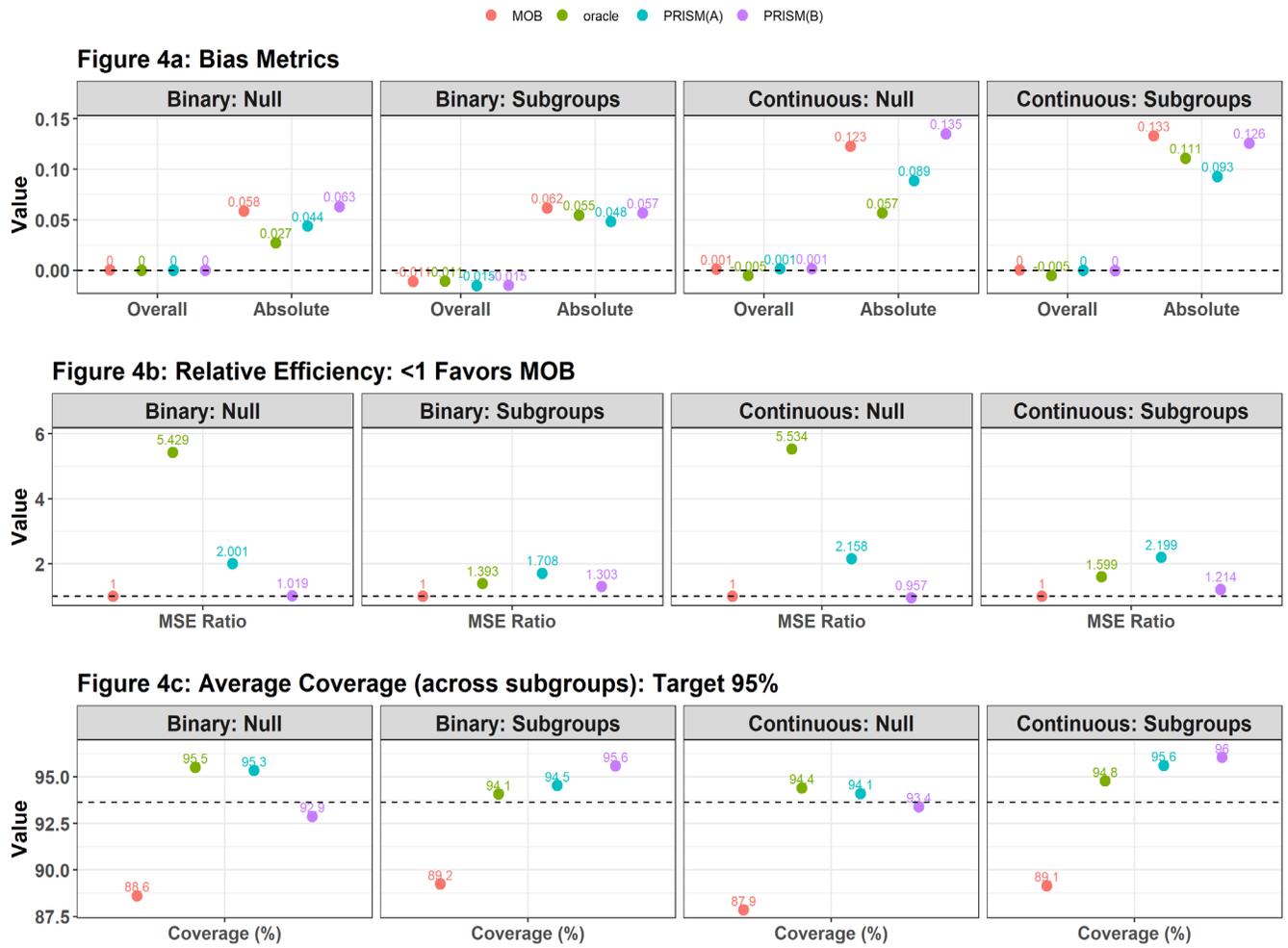

**Note:** This plot illustrates bias, relative efficiency, and coverage properties for three different subgroup methods and the "oracle" which knows the true subgroups. Panels correspond to different simulation scenarios based on outcome type (binary, continuous) and subgroups (Null, Subgroups). PRISM(A) has the lowest bias and compared to PRISM(B) and MOB, the highest efficiency, and valid coverage at the 95% level. The dashed line corresponds to 0.936, which is two standard errors (based on 1000 simulations) below the desired 95% coverage.

## 4.3   Predicted Treatment Assignment

Here, we evaluate how often patients are assigned the "right" treatment when there is in truth a non-zero treatment effect. In the subgroup simulation settings, 70% of patients should receive treatment B while 30% of patients should receive treatment A. In the null treatment setting, all patients should receive treatment B. As a benchmark, we compare PRISM(A) to the "Standard" approach. For a continuous outcome, the "Standard" approach would be to fit a simple OLS/GLM model and if the p-value for the treatment effect is less than 0.05, the decision is to treat all patients with treatment B; otherwise treat all patients with treatment A. For this example, we use an "unadjusted model"; results were similar even when in an "ideal" adjusted analysis where we correctly knew the true predictive covariates, $Y \sim A + X_1 + X_2 + X_3$. For PRISM(A), patients are assigned to treatment B if their subgroup-specific posterior probability (See Section 3.2) exceeds either 0.50 or 0.80; otherwise those patients are assigned to treatment A. See Figure 5.

Classification metrics of interest were accuracy (% of patients correctly assigned), positive predictive value (PPV; among the patients predicted to receive treatment B, what % should have received B?), and negative predictive value (NPV; among the patients predicted to receive A, what % should have received A?). Ideally, all three of these metrics should be high. Starting with the subgroup settings, we see that PRISM(A) yields the highest accuracy, PPV, and NPV. This is true for both 0.50 and 0.80 probability cutoffs. These different probability cutoffs illustrate a general trade-off: lower cutoffs increase NPV while higher cutoffs increase PPV. Lastly, in line with intuition since PRISM(A) uses MOB to find subgroups, using only MOB (with OLS/GLM parameter estimation and Bayesian probabilities) yielded similar classification metrics.

**Figure 5: Predicted Treatment Assignment, PRISM(A) vs Standard Practice**

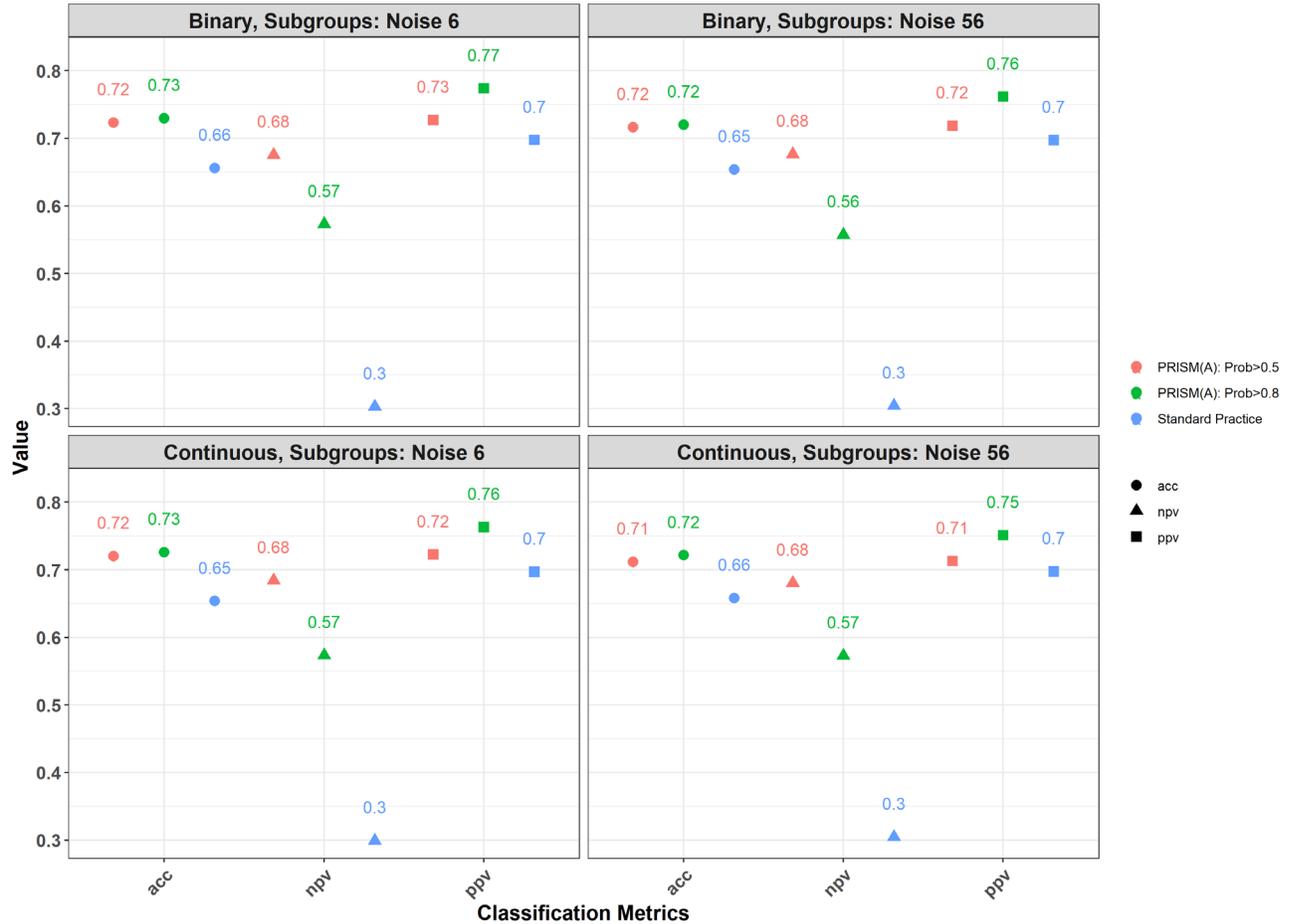

**Note:** This plot illustrates classification metrics for PRISM(A) and "Standard Practice". Standard Practice fits a single treatment-adjusted OLS or GLM model and if the p-value<0.05, all patients will receive treatment "B". Otherwise, patients receive treatment "A". Overall, PRISM(A) yields higher accuracy, higher positive predictive value (ppv), and higher negative predictive value (npv).

## 4.4 Summary of Simulation Results

Overall, these results indicate that (1) Filtering improves the performance of subgroup identification, (2) PRISM(A) tends to select the true predictive variables more often than PRISM(B), (3) PRISM(A) yields the lowest bias, highest relative efficiency, and valid coverage, (4) Compared to "Standard Practice" (treat everyone or not), PRISM(A) yields higher accuracy, PPV, and NPV with regards to assigning the "right" treatment assignment. The final point is especially important as the overall goal of stratified medicine is to assign the "right" treatment to the "right" patient. The first three properties directly relate to the success of treatment assignment, and furthermore, are important since subgroups should ideally be based on the true predictive factors and for any identified subgroup, decision making requires relatively unbiased treatment effect estimates and an understanding of the variability. Lastly, we point out that while in the simulations treatment assignments in PRISM(A) were based on select posterior probability thresholds, similar results can be obtained through using CIs. Depending on the threshold (probability threshold, CI alpha level), there is of course a trade-off between classification metrics that must be considered.

# 5   Return to Clinical Trial Example

Returning to the motivating clinical trial example, we present results using PRISM(A) (See Appendix for details); PRISM(B) and MOB yield similar conclusions. For the overall population the estimated risk difference [95% CI] was    -0.10 [-0.17, -0.03]. After filtering, all variables remained except for CDI severity. See Figure 6 for the main results through a PRISM tree plot. In total, there were four identified subgroups defined by SNP+, prior CDI history, and age. In each terminal node (nodes 3,4,6,7) or identified subgroup, we present the estimated CDI recurrence (%) for placebo and bezlotoxumab, the estimated risk difference (Diff(B-P) [95% CI]), Bayesian probability statements, and the posterior distribution of the estimated risk difference. First, for patients who are SNP- (SNP+ = No), the results vary by prior CDI history (no history, -0.01 [-0.12, 0.10]; history, (-0.11 [-0.30, 0.07]).   For SNP+ patients, the results vary by CDI history and age group (Prior CDI, -0.21 [-0.41, -0.01]; No Prior CDI, age>=65, -0.25 [-0.44, -0.07]; age<65, -0.12 [-0.23, -0.00]). The probability statements are also informative. With the exception of SNP- patients with no prior CDI history, Prob(Diff(B-P)<0) was greater or equal to 89%, suggesting that these subgroups have some benefit from using bezlotoxumab versus placebo. On the other hand, Prob(Diff(B-P)<-0.10) is the estimated probability that a subgroup has an estimated treatment effect beyond the clinically meaningful threshold. In the worst performing subgroup (SNP- with no prior CDI history), Prob(Diff(B-P)<-0.10)=0.05, suggesting that this group (~41% of the population) has limited benefit from using bezlotoxumab. In contrast, Prob(Diff(B-P)<-0.10) was 0.86 and 0.95 respectively for the best performing subgroups (SNP+, Prior CDI; SNP+, No Prior CDI, age>=65).

Overall, these results indicate that patients with no prior CDI history and SNP- status had limited reduction in CDI recurrence from using bezlotoxumab, while patients with prior CDI history or a SNP+ status had a clinically significant reduction in CDI recurrence. In general, the benefiting groups had high baseline risk of CDI recurrence, which is consistent with the regulatory label for bezlotoxumab. Lastly, if we combine the four benefitting groups (SNP+ or SNP- with prior CDI history), the estimated risk difference [95% CI] is -0.17 [-0.25, -0.08].

**Figure 6: PRISM Tree Plot, CDIFF Example**

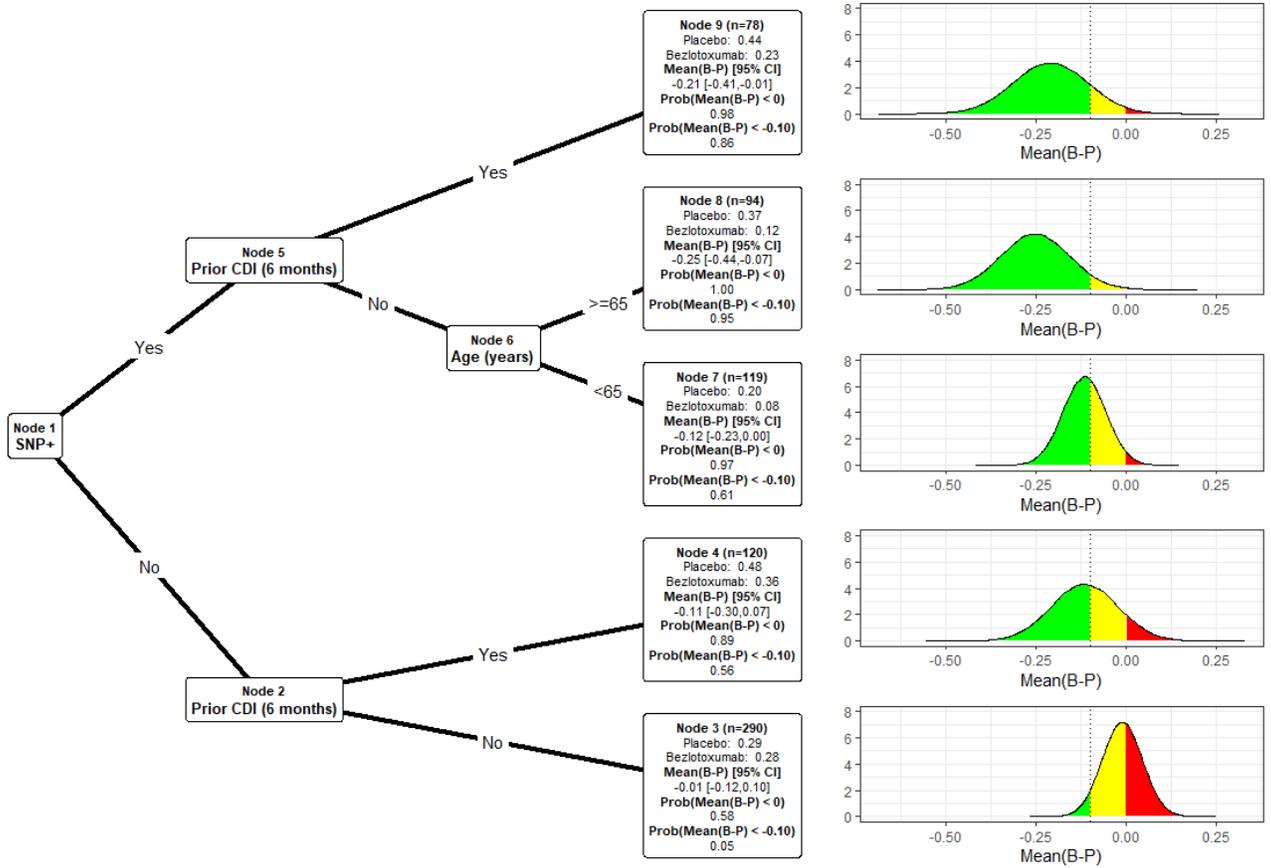

**Note:** In each subgroup, we present the estimated CDI recurrence (%) for placebo/bezlotuxumab, the risk difference [95% CI], the probability that the estimated risk difference is less than the clinically meaningful threshold of -0.10 (obtained through Bayesian approach), and a density plot of the Bayesian posterior distribution (green shade to the left of -0.10; yellow for (-0.10,0); red for >0). Overall, patients who are SNP- with no prior history of CDI do not appear to benefit from bezlotuxumab. Patients who are SNP+ or have prior history of CDI show benefit.

## 6 Discussion

Overall, this paper has described a flexible and powerful subgroup identification framework, PRISM. PRISM is a five-step procedure with the over-arching goal of identifying subgroups and making decisions based on subgroup-specific treatment effects and variability metrics (CIs, SEs, probability statements). While PRISM is flexible with respect to the outcome type and number of treatments, this paper's results focused on a binary treatment scenario (test drug vs control) and binary/continuous endpoints. Based on our simulation study, the best proposed method was PRISM(A), which uses the tree-based MOB to identify subgroups and counterfactual patient-level estimates (PLEs) for decision making. This yielded relatively unbiased subgroup-specific estimates and valid CIs which are crucial for decision making. In our real-data example, PRISM(A) revealed two distinct groups of patients, patients with no previous

CDI history and a SNP- status and patients with prior CDI history or a SNP+ status, with heterogeneous treatment effects.

The goal of stratified medicine is to find subsets or subgroups of patients who are "similar" to each other in terms of treatment-specific responses. For drug development, the challenge is to identify subgroups in a "timely" fashion such that these insights can be used to better design clinical trials and target patients who likely benefit from the test drug. In earlier phases of drug development, due to the limited sample sizes, identifying subgroups or treatment response heterogeneity is more about informing the subsequent phases of drug development. For Phase II/III, if there is substantial evidence that there are subgroups of patients that respond to the test drug and groups of patients that do not respond the decision may be to seek approval for the responder group alone.

The motivation of PRISM was to develop a general-purpose tool for the advancement of stratified medicine. Despite this, while advances in machine learning and development of more robust subgroup models will ultimately improve PRISM, having more patient-level data will have a larger impact. Real world data is one potential solution to expanding the data available for subgroup identification. For example, a recent study using electronic health data replicated associations between tumor mutation burden and response to immunotherapy in non-small lung cancer patients[30]. This illustrated the feasibility of utilizing a real-world data source with clinical and genomic factors for examining differential responses by patient characteristics. Pooling data across clinical trials with study drugs that have similar mechanisms of action, or multiple dose levels, may also help. Modern clinical designs such as the basket trial where the same treatment is studied in multiple indications[31], could also be used to provide more patient-level data to at the very least, understand broad prognostic or risk factors.

In summary, this manuscript developed and illustrated examples using a general-purpose subgroup identification framework, PRISM. Future extensions include a detailed look at survival data, multiple treatments or dose levels, bootstrap/resampling methods, benefit-risk, and development of more robust subgroup models. PRISM (with or without resampling) can also be directly implemented within the "StratifiedMedicine" R Package for continuous, binary, or survival data (experimental) and includes many features not directly discussed here. While there are many challenges, the end goal is to advance stratified medicine and determine whether there are subsets of patients who respond or do not respond to the test drug. Timely identification of subgroups has the potential to accelerate drug development by efficiently allocating resources to more-likely-to-respond patients, expand treatment options for prescribers, and most importantly, improve patient care.

## Appendix

## Non-Parametric Bootstrapping

Generate $b = 1, \ldots, B$ bootstrap resamples and for each resample $b$ with data structure $(\boldsymbol{y_b}, \boldsymbol{a_b}, \boldsymbol{x_b})$, run PRISM (Steps 1-4). In each resample, there are $k_b = 1, \ldots, K_b$ bootstrap subgroups which likely differ from the original $K$ subgroups, both in terms of the rules and the number of identified subgroups. Each bootstrap subgroup $k_b$ will also have an associated point-estimate $\hat{\theta}_{k_b}$. The subgroup-specific estimate in resample $b$ for subgroup $k$, which is identified using the observed data, is then calculated as:

$$\hat{\theta}_{k,b} = \sum_{k_b} n(k \cap k_b) \hat{\theta}_{k_b} \Big/ \sum_{k_b} n(k \cap k_b)$$

where $n(k \cap k_b)$ is the number of observations from the observed data that are in both the original subgroup $k$ and the bootstrap subgroup $k_b$. Next, the smoothed or bagged bootstrap estimate is obtained by averaging the bootstrap estimates:

$$\hat{\theta}_k = \frac{1}{B} \sum_b \hat{\theta}_{k,b}$$

Bootstrap CIs for $\hat{\theta}_k$ are then calculated using the percentile method where the lower/upper limits correspond to the $\alpha/2$ and $1 - \alpha/2$ quantiles of the bootstrap distribution for the subgroup-specific estimates, $(\overline{\theta}_{k,1,\ldots,} \hat{\theta}_{k,b}, \ldots, \hat{\theta}_{k,B})$. Probability statements can also be obtained using the subgroup specific bootstrap distributions. For example, with $I(.)$ denoting an indicator function:

$$\hat{P}(\theta_k > c) = \frac{1}{B} \sum_b I(\hat{\theta}_{k,b} > c)$$

Notably, the bootstrap smoothed estimate yielded similar properties as the PLE estimate (with or without the Bayesian update). The advantage of the bootstrap "smoothed" estimate is that it accounts for the uncertainty of the subgroup model. This approach however requires larger sample sizes. From our experience with both simulations and empirical data-sets, at smaller sample sizes (say N<200) it is more likely that at each bootstrap resample the subgroup model will find no subgroups, thus heavily shrinking the subgroup specific estimates towards the overall mean.

**Simulation Details**

For all simulations, a sample size of N=800 was generated with a 1:1 randomization ratio for the study drug (*A=1*) and standard of care (*A=0*). Covariates were generated from a multivariate normal distribution with mean zero and unit variances. In our simplest setting, there were 12 covariates total ($X_1, \dots, X_{12}$) with three that were prognostic and predictive ($X_1, X_2, X_3$), 3 covariates that were only prognostic ($X_5, X_7, X_{10}$), and 6 noise covariates ($X_4, X_6, X_8, X_9, X_{11}, X_{12}$). The pairwise correlations were set to 0.10 with the following exception: ($X_1, X_5$), ($X_2, X_6$), ($X_3, X_7$) each had pairwise correlations of 0.30 (prognostics moderately correlated with predictive and prognostic covariates). Three of these continuous variables were then discretized into binary: $X_1$=1 if $X_1$>qnorm(0.80) else 0, $X_9$=1 if $X_9$>qnorm(0.70) else 0, and $X_{10}$=1 if $X_{10}$>qnorm(0.40) else 0. To further assess the impact of noise covariates, an additional 50 continuous noise covariates were included in the multivariate normal distribution. These additional noise covariates had a pairwise correlation of 0.10 with all variables. The general point was to make subgroup identification difficult and in fact, we used a similar simulated data-set for an internal subgroup data science contest.

The outcome was continuous with a t-distribution (20 degrees of freedom) and a standard deviation of 0.85 or binomial. The mean of the outcome depended on the prognostic and/or predictive covariates ($X_1, X_2, X_3, X_5, X_7, X_{10}$) and the treatment assignment:

*Continuous Outcome: Generating Model*

$$E(Y) = 1.5 + A * \theta(\boldsymbol{X}) + 0.28X_1 - 0.20 \, std(X_2) + 0.15std(X_3) + 0.14std(X_5) + 0.09std(X_7) + 0.22X_{10}$$

*Binary Outcome: Generating Model*

$$logit(Y) = logit(0.30) + A * \theta(\boldsymbol{X}) + 0.80X_1 - 0.50 \, std(X_2) + 0.40std(X_3) + 0.20std(X_5) + 0.20std(X_7) + 0.30X_{10}$$

where *std(.)* is a standardization function and $\theta(\boldsymbol{X})$ depends on one of two treatment related scenarios: (1) Null setting where there was no treatment difference for all patients, $\theta(\boldsymbol{X}) = 0$,

(2) Subgroup setting where $\theta(X)$ varies and there are four distinct subgroups. In the Subgroup setting, while the overall treatment effect is 0.237 (continuous) or 0.11 (binary), 30% of the patients receive no benefit from the study drug. Subgroups were defined by $(X_1, X_2, X_3)$. See below table for more details. Lastly, we note that the simulation values were chosen such that an unadjusted linear/logistic regression would yield ~90% power in testing the overall treatment effect in the Subgroup settings.

**Supplementary Table 1:** Simulation Treatment Effects by Subgroup and Setting (Continuous)

| Subgroup | Prevalence (%) | E(Y) | Treatment Effect | |
| --- | --- | --- | --- | --- |
| | | | Null Scenario | Subgroup Scenario |
| $X_1 = 1, X_2 < -0.20, X_3 > 0.47$ | 2 | 2.32 | 0 | 0.40 |
| $X_1 = 1, X_2 \geq -0.20, X_3 > 0.47$ | 5 | 2.04 | 0 | 0.40 |
| $X_1 = 1, X_2 < -0.20, X_3 \leq 0.47$ | 5 | 2.02 | 0 | 0.40 |
| $X_1 = 0, X_2 < -0.20, X_3 > 0.47$ | 10 | 1.96 | 0 | 0.40 |
| $X_1 = 1, X_2 \geq -0.20, X_3 \leq 0.47$ | 8 | 1.76 | 0 | 0.33 |
| $X_1 = 0, X_2 \geq -0.20, X_3 > 0.47$ | 15 | 1.68 | 0 | 0.33 |
| $X_1 = 0, X_2 < -0.20, X_3 \leq 0.47$ | 25 | 1.65 | 0 | 0.30 |
| $X_1 = 0, X_2 \geq -0.20, X_3 \leq 0.47$ | 30 | 1.38 | 0 | 0 |
| **Overall** | 100 | 1.66 | 0 | 0.237 |

**Supplementary Table 2:** Simulation Treatment Effects by Subgroup and Setting (Binary)

| Subgroup | Prevalence (%) | E(Y) | Treatment Effect | |
| --- | --- | --- | --- | --- |
| | | | Null Scenario | Subgroup Scenario |
| $X_1 = 1, X_2 < -0.20, X_3 > 0.47$ | 2 | 0.38 | 0 | 0.25 |
| $X_1 = 1, X_2 \geq -0.20, X_3 > 0.47$ | 5 | 0.40 | 0 | 0.25 |
| $X_1 = 1, X_2 < -0.20, X_3 \leq 0.47$ | 5 | 0.35 | 0 | 0.25 |
| $X_1 = 0, X_2 < -0.20, X_3 > 0.47$ | 10 | 0.35 | 0 | 0.25 |
| $X_1 = 1, X_2 \geq -0.20, X_3 \leq 0.47$ | 8 | 0.37 | 0 | 0.17 |
| $X_1 = 0, X_2 \geq -0.20, X_3 > 0.47$ | 15 | 0.36 | 0 | 0.17 |

| | | | |
|---|---|---|---|
| $X_1 = 0, X_2 < -0.20, X_3 \leq 0.47$ | 25 | 0.32 | 0 | 0.11 |
| $X_1 = 0, X_2 \geq -0.20, X_3 \leq 0.47$ | 30 | 0.33 | 0 | 0 |
| **Overall** | 100 | 0.34 | 0 | 0.11 |

For each method with $k = 1, \ldots K$ subgroups identified in each simulation, the following metrics were then calculated:

$$Bias(Overall) = \sum_k n_k (\hat{\theta}_k - \theta_k) / \sum_k n_k$$

$$Bias(Abs) = \sum_k n_k |\hat{\theta}_k - \theta_k| / \sum_k n_k$$

$$MSE = \sum_k n_k (\hat{\theta}_k - \theta_k)^2 / \sum_k n_k$$

$$Coverage = \sum_k n_k I(\theta_k \in CI_{0.95}(\hat{\theta}_k)) / \sum_k n_k$$

where $\theta_k$ is the true treatment effect based on subgroup-specific rules and is estimated using 10,000 simulation patients (not just the N=800 simulated data-set). Note that each metric is weighted by the subgroup-specific sample size. While Bias(Overall) should be roughly zero, it can be misleading as subgroups with large positive bias can be offset by subgroups with large negative bias. Bias(Abs) avoids this issue. $I(\theta_k \in CI_{0.95}(\hat{\theta}_k))$ is an indicator for whether the 95% two-sided CI contains the true treatment effect. Filtering and the subgroup model were also assessed by counting up how many of the predictive and prognostic, and noise variables were included in the subgroup rule definitions. Ideally, the subgroup definitions would always include the true predictive covariates, $(X_1, X_2, X_3)$. All metrics were averaged across 1000 simulations for each scenario.